\documentclass[letterpaper, 10 pt, conference]{ieeeconf}  

\IEEEoverridecommandlockouts                              

\usepackage{cite}
\usepackage{amsmath,amssymb,amsfonts}
\usepackage{algorithmic}
\usepackage{graphicx}
\usepackage{algorithm,algorithmic}
\usepackage{hyperref}
\hypersetup{hidelinks=true}
\usepackage{textcomp}
\usepackage{graphicx}
\usepackage{epsfig}  
\usepackage{amsmath}
\usepackage{indentfirst}  
\usepackage{color}
\usepackage[dvipsnames]{xcolor}
\usepackage{subcaption}
\usepackage{tikz}
\usepackage{soul}
\usepackage{xcolor}
\usepackage{mathtools}
\usepackage{optidef}
\usepackage{breqn}
\usepackage[algo2e,norelsize]{algorithm2e} 

\SetKwInput{kwAdd}{Add}
\SetKwInput{kwSet}{Set}
\SetKwInput{kwReset}{Reset}
\SetKwInput{kwUpdate}{Update}
\SetKwInput{kwClear}{Clear}
\SetKwInput{kwSave}{Save}
\SetKwInput{kwBuild}{Build}
\SetKwInput{kwIni}{Initialization}
\usetikzlibrary{matrix,chains,positioning,decorations.pathreplacing,arrows}

\newcommand{\bG}{{\mathbf G}}

\newcommand{\bX}{{\mathbf X}}

\newcommand{\bL}{{\mathbf L}}

\newcommand{\bI}{{\mathbf I}}

\def\matt#1{\begin{bmatrix}#1\end{bmatrix}}

\def\matt#1{\begin{bmatrix}#1\end{bmatrix}}

\title{\LARGE \bf Distributed Koopman Learning using Partial Trajectories for Control}

\author{Wenjian Hao, Zehui Lu, Devesh Upadhyay, Shaoshuai Mou 
\thanks{This material is based upon work supported by the Defense Advanced Research Projects Agency (DARPA) under of the Learning Introspective Control (LINC) project (grant no. N65236-23-C-8012) Any opinions, ﬁndings and conclusions or recommendations expressed in this material are those of the author(s) and do not necessarily reﬂect the views of the DARPA or the U.S. Government.}
\thanks{W. Hao, Z. Lu, and S. Mou are with the School of Aeronautics and Astronautics, Purdue University, West Lafayette, IN 47907, USA. (Email: \texttt{\{hao93, lu846, mous\}@purdue.edu})}
\thanks{D. Upadhyay is with SAAB Inc. (Email: \texttt{devesh.upadhyay@saabinc.com})}}

\begin{document}

\maketitle
\thispagestyle{empty}
\pagestyle{empty}

\begin{abstract}
This paper proposes a distributed data-driven framework for dynamics learning, termed distributed deep Koopman learning using partial trajectories (DDKL-PT). In this framework, each agent in a multi-agent system is assigned a partial trajectory offline and locally approximates the unknown dynamics using a deep neural network within the Koopman operator framework. By exchanging local estimated dynamics rather than training data, agents achieve consensus on a global dynamics model without sharing their private training trajectories. Simulation studies on a surface vehicle demonstrate that DDKL-PT achieves consensus on the learned dynamics, and each agent attains reasonably small approximation errors on the testing dataset. Furthermore, a model predictive control scheme is developed by integrating the learned Koopman dynamics with known kinematic relations. Results on a reference-tracking task indicate that the distributedly learned dynamics are sufficiently accurate for model-based optimal control.
\end{abstract}

\section{Introduction}
\label{sec:introduction}
Learning system dynamics from system states-inputs data pairs has attracted considerable research attention due to the increasing complexity of the autonomous systems \cite{mamakoukas2021derivative,mezic2015applications, proctor2018generalizing, mauroy2016linear}. In the field of dynamics learning, recent methods include using deep neural networks (DNNs) \cite{murphy2002dynamic, gillespie2018learning}, physics-informed neural networks \cite{raissi2019physics}, and lifting linearization methods such as Koopman operator methods \cite{mezic2015applications, proctor2018generalizing, mauroy2016linear} have been proven to be effective in approximating unknown dynamical systems. 

Among the various approaches, the Koopman operator has gained popularity for its ability to represent known dynamics in a linear form. Two prominent Koopman-based methods, dynamic mode decomposition \cite{schmid2010dynamic} and extended dynamic mode decomposition, have been proposed to lift the state space to a higher-dimensional space, for which the evolution is approximately linear \cite{korda2018linear}. However, selecting lifting functions that ensure the lifted system is linear and accurate remains a significant challenge. To address this problem, various eigen-decomposition-based truncation techniques have been proposed. For instance, \cite{lusch2017data} proposed to use deep learning methods to discover the eigenfunctions of the approximated Koopman operator, and \cite{yeung2019learning, dk2, dk, bevanda2021koopmanizingflows} employed DNNs as lifting functions of the Koopman operator, which are tuned based on collected state-control pairs by minimizing a suitably defined loss function which is also referred as the deep Koopman operator method (DKO). Recent work, such as \cite{hao2024deep}, has extended the DKO method to approximate nonlinear time-varying systems. 

Although Koopman-based methods have been successfully applied to dynamics learning, existing approaches do not directly address scenarios involving large-scale datasets of state–input pairs. To overcome this limitation, we propose a consensus-based framework, termed distributed deep Koopman learning using partial trajectories (DDKL-PT), to approximate nonlinear time-invariant systems (NTIS) with linear models, enabling applications such as optimal control design. Unlike centralized block-wise Koopman operator methods \cite{nandanoori2021data, mukherjee2022learning}, which identify a single global operator for all agents in a multi-agent system (MAS), DDKL-PT assigns each agent an offline partial trajectory and enables local dynamics estimation via a deep neural network embedded in the Koopman operator framework. Through local communication of the estimated dynamics within neighbors, agents reach consensus on a global Koopman representation without sharing their private trajectory data. This distributed approach (i) reduces the computational burden of large-scale learning by distributing the workload across agents, thereby improving scalability, and (ii) preserves data locality and privacy when trajectories are inherently distributed. The main contributions are summarized as follows:
\begin{itemize}
\item We develop a distributed deep Koopman learning algorithm for identifying the dynamics of unknown NTIS, with each agent in a MAS having access only to partial state–input trajectories. The algorithm ensures consensus among agents on the approximated dynamics without sharing private training data during learning.
\item We present a model predictive control (MPC) scheme for a surface vehicle that performs goal-tracking and station-keeping tasks. The controller integrates the learned Koopman dynamics with known kinematic relations, and simulation results demonstrate that the distributedly learned dynamics are sufficiently accurate for model-based optimal control.
\end{itemize}
This paper is structured as follows: In Section \ref{problem_setting}, we introduce the problem. Section \ref{algrithm} details the proposed distributed algorithm. Section \ref{simulation} presents numerical simulations. Finally, Section \ref{conclusion} concludes with a summary of the results.

\textbf{Notations.} Let $\parallel \cdot \parallel$ denote the Euclidean norm. For a matrix $A\in\mathbb{R}^{n\times m}$, $\parallel A \parallel_F$ denotes its Frobenius norm, $A'$ denotes its transpose, and $A^\dagger$ denotes its Moore-Penrose pseudoinverse. $\bI_{n}$ denotes a ${n \times n}$ identity matrix.

\section{The Problem}\label{problem_setting}
Consider the following discrete-time nonlinear time-invariant systems (NTIS): \begin{equation}\label{eq_unknowndyn}
    \boldsymbol{x}(t+1) = \boldsymbol{f}(\boldsymbol{x}(t), \boldsymbol{u}(t)),
\end{equation}
where $t=0,1,2,\cdots$ denotes the time index, $\boldsymbol{x}(t)\in \mathbb{R}^n$ and $\boldsymbol{u}(t) \in \mathbb{R}^m$ denote the system state and control input at time $t$, respectively, and $\boldsymbol{f}: \mathbb{R}^n\times\mathbb{R}^m\rightarrow\mathbb{R}^n$ is the time-invariant mapping function which is assumed to be unknown.
A trajectory of system states-inputs from time $0$ to time $T$ is represented by: 
\begin{equation}\label{eq_traj}
    \boldsymbol{\xi} = \{(\boldsymbol{x}_t, \boldsymbol{u}_t): t = 0, 1,2, \cdots, T\},
\end{equation} where $(\boldsymbol{x}_t, \boldsymbol{u}_t)$ denotes the observed fixed state–input pair, which is distinguished from the state-input variables $(\boldsymbol{x}(t),\boldsymbol{u}(t))$ used elsewhere in this paper.

The goal of the centralized DKO is to, given the entire trajectory $\boldsymbol{\xi}$, identify constant matrices $A^*\in\mathbb{R}^{r\times r}$, $B^*\in\mathbb{R}^{r\times m}$, $C^*\in\mathbb{R}^{n\times r}$, and a parameter vector $\boldsymbol{\theta}^*\in\mathbb{R}^p$ such that for any $0\leq t\leq T-1$, the following holds:
\begin{equation}\label{eq_DKO}
\begin{aligned}
\boldsymbol{g}(\boldsymbol{x}_{t+1},\boldsymbol{\theta}^*) &= A^* \boldsymbol{g}(\boldsymbol{x}_t,\boldsymbol{\theta}^*) + B^* \boldsymbol{u}_t,\\ 
    \boldsymbol{x}_{t+1} &= C^*\boldsymbol{g}(\boldsymbol{x}_{t+1},\boldsymbol{\theta}^*),
\end{aligned}
\end{equation} where $\boldsymbol{g}(\cdot,\boldsymbol{\theta}^*):\mathbb{R}^n\rightarrow\mathbb{R}^r$ is a lifting function with known structure and lifting dimension $r\geq n$. The first equation represents linear dynamics in the lifted space. In contrast, the second equation assumes a linear mapping between the lifted state $\boldsymbol{g}(\boldsymbol{x}_t, \boldsymbol{\theta}^*)$ and its original state $\boldsymbol{x}_t$. 

\begin{figure}
    \centering
    \includegraphics[width=0.49\textwidth]{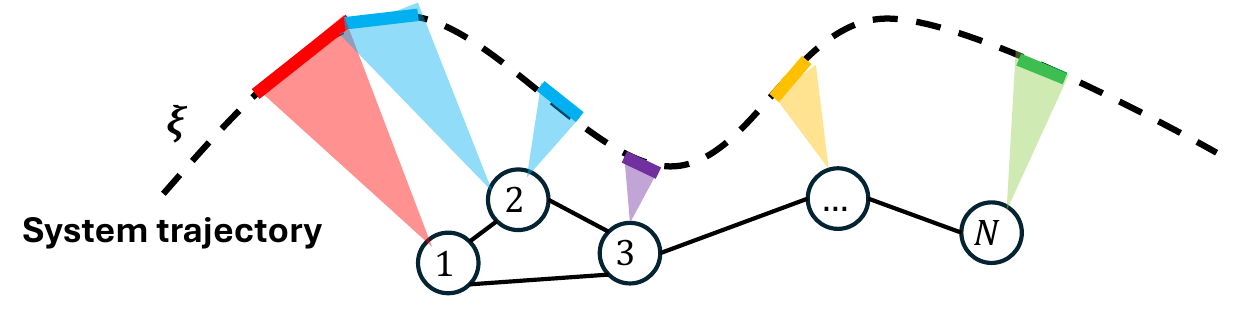}
    \caption{Illustration of the distributed Koopman learning, where each agent is only available for a partial trajectory.}
    \label{fig:illus}
\end{figure}

Consider a group of $N \geq 1$ agents with its node set denoted as $\mathcal{V} = \{1,2,\dots, N\}$. Each agent $i \in \mathcal{V}$ can receive/send information from/to its neighbor set $\mathcal{N}_i \subset \mathcal{V}$, $i \in \mathcal{N}_i$. In other words, the communication between agent $i$ and agent $i$'s neighbors in $\mathcal{N}_i$ is bidirectional.
Neighbor relations between distinct pairs of agents are characterized by a self-arced undirected graph $\mathbb{G} = \{\mathcal{V},\mathcal{E}\}$ such that $(i,j)\in\mathcal{E}$ if and only if agents $i$ and $j$ are neighbors. We assume $\mathbb{G}$ is connected. As shown in Fig. \ref{fig:illus}, we assume a MAS of which any agent $i$ observes partial trajectory of \eqref{eq_traj} denoted by 
\begin{equation}\label{eq_batch}
    \boldsymbol{\xi}_i=\{(\boldsymbol{x}_{t_{i,k}},\boldsymbol{u}_{t_{i,k}}): k = 0,1,2,\cdots,T_i\},
\end{equation}
where $t_{i,0}$ and $t_{i,T_i}$ denotes the starting time and end time of trajectory $\boldsymbol{\xi}_i$ respectively with $0\leq t_{i,0} < t_{i,T_i}\leq T$. Note that no restrictions are imposed on $\boldsymbol{\xi}_i$, that is, it may contain as few as two data pairs, and trajectory segments across different agents may overlap.

When the dataset $\boldsymbol{\xi}$ contains a large number of state–input pairs (i.e., $T$ is large), a centralized DKO learner may face significant difficulties in efficiently processing the data. Moreover, the partially observed trajectory $\boldsymbol{\xi}_i$ available to each agent is generally insufficient for independently identifying the dynamics in \eqref{eq_DKO}. Motivated by this limitation, the \textbf{problem of interest} is to obtain \eqref{eq_DKO} by developing a distributed DKO framework, wherein each agent $i$ maintains its own estimated dynamics set \begin{equation}\label{eq_ddkr}
    \mathcal{K}_i = \{A_i, B_i, C_i, \boldsymbol{\theta}_i\},
\end{equation} which serves as a local approximation of $\{A^*, B^*, C^*, \boldsymbol{\theta}^*\}$ in \eqref{eq_DKO}. Agents iteratively update their estimates $\mathcal{K}_i$ via both local partial trajectory and collaboratively exchanging $\mathcal{K}_i$ with their neighbors.

\section{The Algorithm}\label{algrithm}
In this section, we outline the main challenges and key ideas underlying the proposed problem, followed by a proposed distributed algorithm for achieving dynamics in \eqref{eq_DKO}.
\subsection{Challenges and Key Ideas}
The main challenge in achieving \eqref{eq_DKO} lies in the fact that each agent $i$ in the MAS only observes partial trajectories of \eqref{eq_traj}, which are insufficient to recover \eqref{eq_DKO} for all state-input pairs in \eqref{eq_traj}. To overcome this limitation, we propose a distributed algorithm in which each agent learns its local dynamics from available training data and subsequently exchanges the learned dynamics with its neighbors to estimate \eqref{eq_DKO}. This procedure can be formulated as a multi-agent optimization problem.

\textbf{Local dynamics estimation.} Suppose a MAS of $N$ agents, where each agent $i$ can only access a set of trajectory segments $\boldsymbol{\xi}_i$ from \eqref{eq_batch}. Following Koopman operator theory \cite{koopman1932dynamical}, to approximate the unknown dynamics in \eqref{eq_unknowndyn}, each $i$ has to identify constant matrices $A_i^*\in\mathbb{R}^{r\times r}$, $B_i^*\in\mathbb{R}^{r\times m}$, $C_i^*\in\mathbb{R}^{n\times r}$, and a parameter vector $\boldsymbol{\theta}_i^*\in\mathbb{R}^p$ such that the following holds: \begin{align}
        \boldsymbol{g}(\boldsymbol{x}_{t_{i, k+1}},\boldsymbol{\theta}_i^*) &= A_i^* \boldsymbol{g}(\boldsymbol{x}_{t_{i, k}},\boldsymbol{\theta}_i^*) + B_i^* \boldsymbol{u}_{t_{i, k}} \label{eq_dis_lift},  \\ \boldsymbol{x}_{t_{i, k+1}} &= C_i^*\boldsymbol{g}(\boldsymbol{x}_{t_{i, k+1}},\boldsymbol{\theta}_i^*) \label{eq_dis_map}, 
\end{align}
To this end, we define for agent $i$ the local dynamics learning error function over its partially observed trajectory as
\begin{equation}\label{eq_local_error}
\begin{aligned}
&\mathbf{L}_i(A_i, B_i, C_i, \theta_i) = 
\frac{1}{2 T_i}\sum_{k=0}^{T_i-1} || \boldsymbol{x}_{t_{i,k}} - C_i \boldsymbol{g}(\boldsymbol{x}_{t_{i,k}},\boldsymbol{\theta}_i)||^2 \\
& \  + || \boldsymbol{g}(\boldsymbol{x}_{t_{i,k+1}},\boldsymbol{\theta}_i) - A_i\boldsymbol{g}(\boldsymbol{x}_{t_{i,k}},\boldsymbol{\theta}_i) - B_i \boldsymbol{u}_{t_{i,k}} ||^2,
\end{aligned}
\end{equation} where the first and second terms are designed to approximate \eqref{eq_dis_map} and \eqref{eq_dis_lift}, respectively. Furthermore, we construct the following data matrices from $\boldsymbol{\xi}_i$:
\begin{equation}\label{xyudata}
    \begin{aligned}
    \bX_i &= \matt{\boldsymbol{x}_{t_{i,0}}, \boldsymbol{x}_{t_{i,1}}, \cdots, \boldsymbol{x}_{t_{i,T_i-1}}} \in \mathbb{R}^{n \times T_i},\\
    \mathbf{U}_i &= \matt{\boldsymbol{u}_{t_{i,0}}, \boldsymbol{u}_{t_{i,1}}, \cdots, \boldsymbol{u}_{t_{i,T_i-1}}} \in \mathbb{R}^{m \times T_i}, \\ 
    \bG_i &= \matt{\boldsymbol{g}(\boldsymbol{x}_{t_{i,0}},\boldsymbol{\theta}_i), \cdots, \boldsymbol{g}(\boldsymbol{x}_{t_{i,T_i-1}},\boldsymbol{\theta}_i)} \in \mathbb{R}^{r \times T_i},\\
    \mathbf{\bar G}_i &= \matt{\boldsymbol{g}(\boldsymbol{x}_{t_{i,1}},\boldsymbol{\theta}_i), \cdots, \boldsymbol{g}(\boldsymbol{x}_{t_{i,T_i}},\boldsymbol{\theta}_i)}\in \mathbb{R}^{r \times T_i}.
    \end{aligned}
\end{equation}
This allows \eqref{eq_local_error} to be rewritten in the following compact form:
\begin{equation}\label{eq_local_error_final}
\begin{aligned}
\mathbf{L}_i(A_i, B_i, C_i, \boldsymbol{\theta}_i) = \frac{1}{2T_i} (\parallel \mathbf{\bar G}_i - \matt{A_i\ B_i} \begin{bmatrix}
\bG_i \\ \mathbf{U}_i
\end{bmatrix}\parallel_F^2 \\+ \parallel \bX_i - C_i\bG_i \parallel_F^2).
    \end{aligned}
\end{equation}
\textbf{Multi-agent optimization problem.} Based on \eqref{eq_local_error_final}, we formulate the following multi-agent optimization problem to approximate \eqref{eq_DKO}:
\begin{equation}\label{eq_obj}
\begin{aligned}
\min_{\{A_i, B_i, C_i, \boldsymbol{\theta}_i\}_{i=1}^N}&\sum_{i=1}^N \mathbf{L}_i(A_i, B_i, C_i, \boldsymbol{\theta}_i)\\
\text{subject to}\quad  A_1 &= A_2 = \cdots = A_N,\\
B_1 &= B_2 = \cdots = B_N,\\
C_1 &= C_2 = \cdots = C_N,\\
\boldsymbol{\theta}_1 &= \boldsymbol{\theta}_2 = \cdots = \boldsymbol{\theta}_N.
\end{aligned}
\end{equation}

\subsection{Algorithm}
We now propose a distributed algorithm to solve \eqref{eq_obj}, which proceeds in two steps. First, for a fixed parameter vector $\boldsymbol{\theta}_i$, problem \eqref{eq_obj} is minimized with respect to the dynamics matrices $A_i$, $B_i$, and $C_i$. Second, given the resulting fixed dynamics matrices from the first step, problem \eqref{eq_obj} is further minimized with respect to $\boldsymbol{\theta}_i$, as detailed below.

\textbf{Step 1: Distributed Learning of Dynamics Matrices.} While existing distributed optimization methods \cite{nedic2010constrained, qu2017harnessing} can be applied, they generally require all agents to adopt a common step size. To remove this restriction while preserving exponential convergence, we adapt the algorithm in \cite{wang2019distributed} and derive the following distributed update scheme.

Let $c$ denote a user-specified positive constant. The weights satisfy $w_{ij} > 0$ if $j\in\mathcal{N}_i$ and $w_{ij} = 0$ otherwise, with $d_i = \sum_{j\in\mathcal{N}_i} w_{ij}$. The weights are assumed to be symmetric, i.e., $w_{ij} = w_{ji}$. All agents are initialized with a common parameter, i.e., $\boldsymbol{\theta}_1=\boldsymbol{\theta}_2=\cdots=\boldsymbol{\theta}_N$. Let $M_i = [A_i, B_i]$, $N_i = [\bG_i', \mathbf{U}_i']'$, and $s$ denotes the iteration index. The proposed distributed update rule for $M_i$ and $C_i$ is given as follows:
\begin{equation}\label{eq_Mmatupdate}
\small{
    \begin{bmatrix}
        M_i'(s+1) \\ E_i(s+1)
    \end{bmatrix} = F_i^{-1} \begin{bmatrix}
        d_i M_i'(s) + \sum_{j\in\mathcal{N}_i} w_{ij}E_j(s) + cN_i\mathbf{\bar G}_i'\\
        - \sum_{j\in\mathcal{N}_i} w_{ij}M_j'(s) + d_i E_i(s)
    \end{bmatrix}}
\end{equation}
and
\begin{equation}\label{eq_Cmatupdate}
\small{
    \begin{bmatrix}
        C_i'(s+1) \\ \hat{E}_i(s+1)
    \end{bmatrix} = \hat{F}_i^{-1} \begin{bmatrix}
        d_i C'_i(s) + \sum_{j\in\mathcal{N}_i} w_{ij}\hat{E}_j(s) + c\mathbf{G}_i\mathbf{X}_i'\\
        - \sum_{j\in\mathcal{N}_i} w_{ij}C_j'(s) + d_i \hat{E}_i(s),
    \end{bmatrix}},
\end{equation}
where $E_i\in\mathbb{R}^{r\times (r+m)}$ and $\hat{E}_i\in\mathbb{R}^{n\times r}$ are auxiliary matrices associated with agent $i$ initialized from arbitrary values $E_i(0)$ and $\hat{E}_i(0)$. The matrices $F_i$ and $\hat{F}_i$ are given by
\[\small F_i = \begin{bmatrix}
    d_i \bI_n + cN_iN_i' &d_i \bI_n \\ -d_i \bI_n  &d_i \bI_n 
\end{bmatrix}, \hat{F}_i = \begin{bmatrix}
    d_i \bI_n + c\mathbf{X}_i\mathbf{X}_i' &d_i \bI_n \\  -d_i \bI_n  &d_i \bI_n 
\end{bmatrix}.\] 
As shown in \cite{wang2019distributed}, for fixed $\boldsymbol{\theta}_i$, the update rules \eqref{eq_Mmatupdate}-\eqref{eq_Cmatupdate} converge exponentially fast to the optimal solution of \eqref{eq_obj}.

\textbf{Step 2: Distributed Parameter Tuning.} Suppose that $M_i$ and $C_i$ in \eqref{eq_local_error_final} are constant matrices obtained from Step $1$, the loss function in \eqref{eq_local_error_final} is then only determined by $\boldsymbol{\theta}_i$ as:
\begin{equation}
\begin{aligned}
\mathbf{L}_i(\boldsymbol{\theta}_i) = \frac{1}{2T_i} (\parallel \mathbf{\bar G}_i - M_i \begin{bmatrix}
\bG_i \\\mathbf{U}_i
\end{bmatrix}\parallel_F^2 + \parallel \bX_i - C_i \bG_i \parallel_F^2). \nonumber
\end{aligned}
\end{equation}
The parameter $\boldsymbol{\theta}_i$ is updated following the distributed subgradient methods for multi-agent optimization in \cite{nedic2009distributed}:
\begin{equation}\label{eq_dis_gd}
    \boldsymbol{\theta}_i(s+1) = \sum_{j\in\mathcal{N}_i}\hat{w}_{ij}\boldsymbol{\theta}_j(s) - \alpha_i(s)\nabla_{\boldsymbol{\theta}_i}\bL_i(\boldsymbol{\theta}_i(s)),
\end{equation}
where $\alpha_i(s)$ denotes the diminishing learning rate of agent $i$ at iteration $s$. The weights satisfy $\hat{w}_{ij}>0$ if $j\in\mathcal{N}_i$, and $\hat{w}_{ij} =0$ otherwise. Let $\hat{W}\in\mathbb{R}^{N\times N}$ denote the weight matrix with its $ij$-th entry $\hat{w}_{ij}$, we assume $\hat{W}$ is doubly stochastic.

In summary, we have the following algorithm referred to as \emph{a distributed deep Koopman learning using partial trajectories} (DDKL-PT) in the rest of this paper.
\begin{algorithm}
\caption{Distributed Deep Koopman Learning using Partial Trajectories}\label{algorithm:DDKL-PT}
\begin{algorithmic}[1]
 \renewcommand{\algorithmicrequire}{\textbf{Input:}}
 \renewcommand{\algorithmicensure}{\textbf{Initialization:}}
 \REQUIRE Each agent observes partial trajectory $\xi_i$ in \eqref{eq_batch}.
 \ENSURE  Each agent $i$ constructs $\boldsymbol{g}(\cdot, \boldsymbol{\theta}_i): \mathbb{R}^n \rightarrow \mathbb{R}^r$ with nonzero $\boldsymbol{\theta}_i(0)\in\mathbb{R}^p$, where $\boldsymbol{\theta}_1(0) = \boldsymbol{\theta}_2(0) = \cdots = \boldsymbol{\theta}_N(0)$.
Each agent $i$ initializes random matrices $M_i(0)\in\mathbb{R}^{r\times (r+m)}$, $C_i(0)\in\mathbb{R}^{n\times r}$, $E_i(0)\in\mathbb{R}^{r\times (r+m)}$, $\hat{E}_i(0)\in\mathbb{R}^{n\times r}$, sets the iteration steps $S$ and $\bar{S}$,
the learning rates $\{\alpha_i(s)\}_{s=0}^{\bar{S}}$, and weights $w_{ij}$, $c$ and $\hat{w}_{ij}$. 
  \FOR {$s\leftarrow 0$ to $S$}
  \STATE Each agent $i$ updates $M_i(s)$ and $C_i(s)$ following \eqref{eq_Mmatupdate} and \eqref{eq_Cmatupdate}, respectively. 
  \ENDFOR
  \FOR {$s\leftarrow 0$ to $\bar{S}$}
  \STATE
Each agent $i$ updates $\boldsymbol{\theta}_i(s)$ following \eqref{eq_dis_gd}, where $\nabla_{\boldsymbol{\theta}_i}\bL_i(\boldsymbol{\theta}_i(s))$ is computed using $M_i(S)$ and $C_i(S)$.
  \ENDFOR
  \STATE Save the resulting $\{A_i(S), B_i(S), C_i(S), \boldsymbol{\theta}_i(\bar{S})\}$.
 \end{algorithmic} 
 \end{algorithm}

\section{Simulations}\label{simulation}
\begin{figure}[ht]
    \centering
    \includegraphics[width=0.38\textwidth]{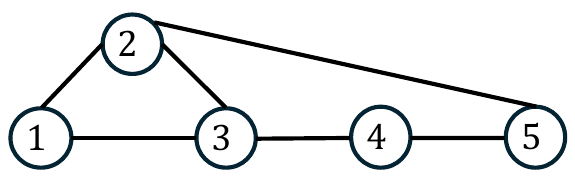}
    \caption{Five-agent connected network with self-arcs omitted for clarity.}
    \label{fig:MAS}
\end{figure}

In this section, we first evaluate the dynamics learning performance of the proposed DDKL-PT algorithm by analyzing its prediction error on a testing dataset. We then present a motivating example that demonstrates the use of DDKL-PT in model-predictive control (MPC) design for a surface vehicle, with each agent in the MAS performing goal-tracking and station-keeping tasks. All simulations consider a five-agent MAS, depicted in Fig. \ref{fig:MAS} (self-arcs omitted for clarity), where each agent observes only a partial trajectory $\boldsymbol{\xi}_i$ as defined in \eqref{eq_batch}.

We first consider the surface vehicle dynamics from \cite{bingham2019toward}, characterized by a $6$-dimensional state vector $\boldsymbol{x}(t) = \matt{\boldsymbol{p}(t)^\prime, \boldsymbol{v}(t)^\prime}^\prime \in\mathbb{R}^6$, where $\boldsymbol{p}(t) = [p_{\mathrm{x}}(t), p_{\mathrm{y}}(t), \phi(t)]^\prime$ represents the 
$x$ and $y$ positions and yaw angle of the surface vehicle in global coordinates, and $\boldsymbol{v}(t) = \matt{v_{\mathrm{x}}(t), v_{\mathrm{y}}(t), \dot{\phi}(t)}^\prime$ denotes the corresponding velocities in local coordinates. The system is controlled by a $2$-dimensional control input $\boldsymbol{u}(t) = \matt{u_{\mathrm{left}}, u_{\mathrm{right}}}^\prime$, where $-1 \leq u_{\mathrm{left}}\leq 1$ and $-1 \leq u_{\mathrm{right}}\leq 1$ represent the thrusts of the left and right motors, respectively. 

A nominal trajectory $\boldsymbol{\xi}$ over the time interval $0\leq t\leq 5000$ is generated by driving the surface vehicle with control inputs randomly sampled from a normal distribution. Partial observations for each agent are collected over the intervals $0\leq t_{1,k} \leq 600$, $600 \leq t_{2,k} \leq 1600$, $1600 \leq t_{3,k} \leq 2800$, $2800 \leq t_{4,k} \leq 3600$, and $3600 \leq t_{5,k} \leq  4000$. The remaining segment, $4000\leq t\leq 5000$, is reserved for testing the learned models.

\subsection{Dynamics Learning Evaluation}\label{sec_dl}
The objective is to learn the mapping from velocity $\boldsymbol{v}_t$ and control input $\boldsymbol{u}_t$ to the next velocity $\boldsymbol{v}_{t+1}$ for all agents in the MAS. For implementation of DDKL-PT, we set $w_{ij}=1$ and $c=1$ for the update rules \eqref{eq_Mmatupdate}-\eqref{eq_Cmatupdate}. The weights $\hat{w}_{ij}$ in \eqref{eq_dis_gd} are chosen as uniform neighbor weights, i.e., each agent assigns equal weights to its neighbors. The initial DNNs $\boldsymbol{g}(\boldsymbol{v}_t, \boldsymbol{\theta}_i):\mathbb{R}^3\rightarrow\mathbb{R}^8$ ($r=8$) are constructed with two hidden layers of $256$ nodes and $ReLU$ activation functions for all agents. The proposed method is compared with two baselines: \begin{itemize}
    \item Centralized DKO: Approximates \eqref{eq_DKO} following \cite{hao2023optimal} using the full trajectory $0\leq t\leq 4000$, with the same initialization $\boldsymbol{g}(\cdot,\boldsymbol{\theta}_i)$ as used in the proposed method.
    \item Multilayer perceptron (MLP): Learns a mapping $\boldsymbol{g}(\boldsymbol{v}_k,\boldsymbol{u}_k,\boldsymbol{\hat\theta}):\mathbb{R}^3\times\mathbb{R}^2\rightarrow\mathbb{R}^3$, where $\boldsymbol{g}(\cdot,\cdot,\boldsymbol{\hat\theta})$ consists of two hidden layers with $256$ nodes and $ReLU$ activation functions. The optimal parameters $\boldsymbol{\hat\theta}$ are obtained by minimizing \[\bL(\boldsymbol{\hat\theta})=\frac{1}{4000}\sum_{k=0}^{3999}\|\boldsymbol{g}(\boldsymbol{v}_k,\boldsymbol{u}_k,\boldsymbol{\hat\theta}) - \boldsymbol{v}_{k+1}\|^2.\] 
\end{itemize} For a fair comparison, all methods use the same training dataset ($0\leq t\leq 4000$) and testing dataset ($4000\leq t\leq 5000$), a constant learning rate of $10^{-4}$, the Adam optimizer \cite{kingma2014adam}, and the same terminal training accuracy of $7\times 10^{-6}$ for all methods. To reduce the effect of randomness in DNN training, each experiment is repeated $10$ times.

\textbf{Evaluation Metrics.} We first demonstrate the consensus achieved by the proposed method during training by measuring the differences between the dynamics matrices and lifting DNNs learned by individual agents in the MAS and those obtained by the centralized DKO method. We then evaluate the learned models on the test dataset and compare their estimation errors with those of benchmark methods. The evaluation metric on the testing dataset is defined as \[V = \frac{1}{ 10^4}\sum_{j=1}^{10}\sum_{k=4000}^{4999}\|\boldsymbol{\hat{v}}_{k+1}^j - \boldsymbol{v}_{k+1}\|^2.\] Here, $\boldsymbol{\hat{v}}_{k+1}^j$ denotes the one-step prediction generated from $(\boldsymbol{v}_k^j, \boldsymbol{u}_k^j)$ in the $j$-th run. For the proposed DDKL-PT method, $\boldsymbol{\hat{v}}_{k+1}^j$ is obtained by averaging the predictions of all agents in the MAS, i.e., $\boldsymbol{\hat{v}}_{k+1}^j = \frac{1}{5}\sum_{i=1}^5\boldsymbol{\hat{v}}_{i, k+1}^j$.

\textbf{Results analysis.} \begin{figure}
\centering
\begin{subfigure}[b]{0.46\textwidth}
\centering
\includegraphics[width=\textwidth]{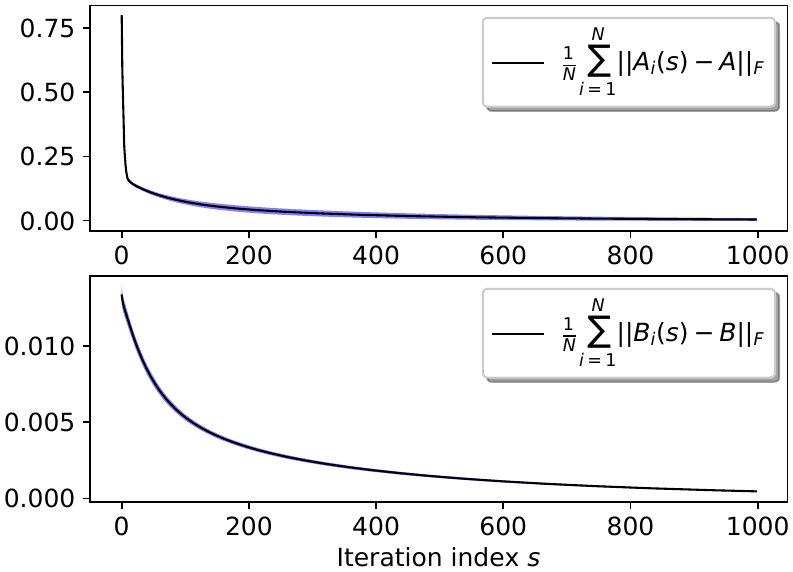}
\caption[]%
{Consensus of $A_i$ and $B_i$ matrices.}    
\label{fig:mat_cons}
\end{subfigure}
\begin{subfigure}[b]{0.48\textwidth}
\centering
\includegraphics[width=\textwidth]{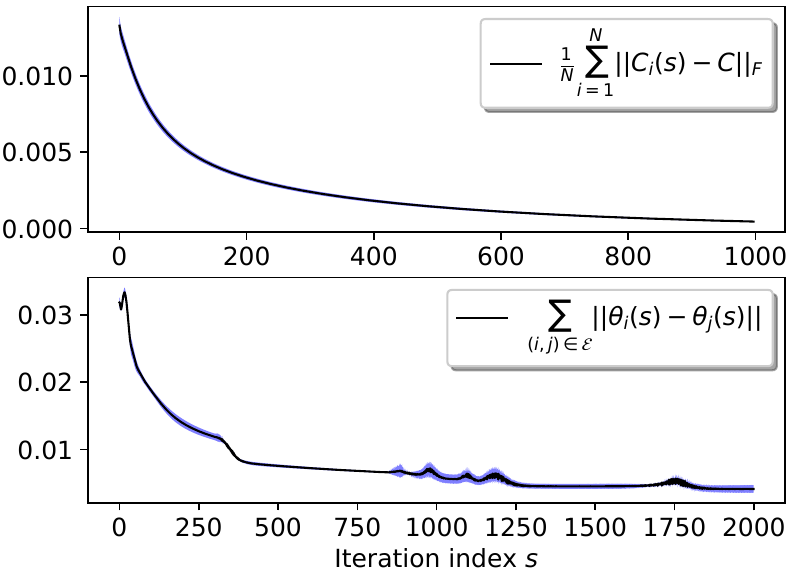}
\caption[]%
{Consensus of $C_i$ matrix and $\boldsymbol{\theta}_i$.}
\label{fig:theta_cons}
\end{subfigure}
\caption{Consensus of the learned dynamics of each agent $i$ during the learning process, where the solid black line denotes the average of the $10$ experiments, and the blue shadow denotes the standard deviation.} 
\label{fig:consensus}
\end{figure} 
\begin{center}
\begin{tabular}{ |c|c|c|c|c|c| }
\hline
 Methods & Estimation errors on testing dataset \\
\hline
{$V_{\textrm{DDKL-PT}}$} & 0.0284$\pm$0.0041 \\
\hline
{$V_{\textrm{DKO}}$} & 0.0179$\pm$0.0016 \\
\hline {$V_{\textrm{MLP}}$} & 0.0205$\pm$0.0028 \\ \hline
\end{tabular}
\captionof{table}{Average estimation errors and standard deviations computed over the testing data across $10$ experimental runs.}
\label{test_err}
\end{center}
Figs. \ref{fig:mat_cons}-\ref{fig:theta_cons} illustrate that the matrices $A_i, B_i, C_i$ obtained by each agent through \eqref{eq_Mmatupdate}-\eqref{eq_Cmatupdate} converge to the dynamics matrices $A, B, C$ obtained from the centralized DKO using the complete trajectory and the same initial $\boldsymbol{g}(\cdot,\boldsymbol{\theta})$. Similarly, the tunable parameters $\boldsymbol{\theta}_i$ reach consensus via the update rule \eqref{eq_dis_gd} during the training process.

Furthermore, to evaluate the model performance on the testing dataset, a one-way analysis of variance (ANOVA) \cite[Chapter~14]{lowry2014concepts} was performed to compare testing errors among the three methods: $V_{\textrm{DDKL-PT}}$ (proposed, distributed), $V_{\textrm{DKO}}$ (centralized), and $V_{\textrm{MLP}}$ (centralized). The results reveal a statistically significant difference in errors between at least two groups, with a p-value of $0.05$. Tukey’s HSD test further establishes the ranking DKO $<$ MLP $<$ DDKL-PT. 
Table~\ref{test_err} summarizes the testing errors. The higher error of the proposed DDKL-PT method arises from the distributed nature of the dataset across agents, in contrast to the centralized training of DKO and MLP. Nevertheless, the simulations in Section~\ref{subsec:mpc} demonstrate that the Koopman representations obtained by DDKL-PT provide sufficient accuracy for the optimal control tasks. 


\subsection{MPC Design with Learned DDKL-PT Dynamics} \label{subsec:mpc}
Consider any agent $i$ in Fig. \ref{fig:MAS}. An MPC controller is designed using the learned dynamics obtained from the proposed DDKL-PT framework to drive the surface vehicle from the initial condition $x(0) = \matt{20, 10, \pi/3, 0, 0, 0}^\prime$ to goal state $x_{\mathrm{goal}} = \matt{0, 0, \pi/2, 0, 0, 0}^\prime$. 

\textbf{Combined Dynamics of DKO and Kinematics:} The learned deep Koopman dynamics are expressed as
\begin{equation}\label{eq_boat_dko}
    \boldsymbol{v}(t+1)=C_i^*\Big(A_i^* \boldsymbol{g}(\boldsymbol{v}(t),\boldsymbol{\theta}_i^*) + B_i^* \boldsymbol{u}(t)\Big) = \begin{bmatrix}
    v_{\mathrm{x}}(t+1)\\ v_{\mathrm{y}}(t+1) \\ \dot{\phi}(t+1)
\end{bmatrix},
\end{equation} where $A_i^*$, $B_i^*$, $C_i^*$, and $\boldsymbol{\theta}_i^*$ come from Section~\ref{sec_dl}. This model predicts the surface vehicle's future velocities in local coordinates.

To obtain full-state predictions, the learned Koopman model \eqref{eq_boat_dko} is integrated with the vehicle’s kinematic dynamics: \begin{equation}\label{eq_boat_kin}
    \boldsymbol{p}(t+1) = \boldsymbol{p}(t) + \boldsymbol{\dot{p}}(t)\Delta t, \quad \Delta t = 0.02s,
\end{equation} where $\boldsymbol{\dot{p}}(t)$ is computed based on the DKO-predicted velocities from  \eqref{eq_boat_dko} as
\begin{equation}
\boldsymbol{\dot{p}}(t) = \begin{bmatrix}
    \cos(\phi(t))v_{\mathrm{x}}(t+1) - \sin(\phi(t))v_{\mathrm{y}}(t+1) \\
    \sin(\phi(t))v_{\mathrm{x}}(t+1) + \cos(\phi(t))v_{\mathrm{y}}(t+1) \\ \dot{\phi}(t+1)
\end{bmatrix}. \nonumber
\end{equation} The combined dynamics are therefore defined as
\begin{equation}\label{eq_combined_dyn}
    \boldsymbol{x}(t+1) = \boldsymbol{F}(\boldsymbol{x}(t), \boldsymbol{u}(t)) =  \begin{bmatrix}
        \boldsymbol{p}(t+1) \\ \boldsymbol{v}(t+1)
    \end{bmatrix},
\end{equation} where $\boldsymbol{p}(t+1)$ and $\boldsymbol{v}(t+1)$ are given in \eqref{eq_boat_kin} and \eqref{eq_boat_dko}, respectively.

\textbf{MPC Setup.} At each time step $t=0,1,2,\cdots$ the following MPC problem is solved for agent $i$:
\begin{equation}
    \begin{aligned}
        \min_{\boldsymbol{u}(0),\cdots,\boldsymbol{u}(K-1)} \sum_{t=0}^{K-1}c_t + c_f, \quad
    \text{s.t.} \quad 
    \eqref{eq_combined_dyn},\ u(t) \in \mathcal{U}, \nonumber
    \end{aligned}
\end{equation}
with horizon length $K=30$. The stage cost is defined as \[c_t = (\boldsymbol{x}(t)-\boldsymbol{x}_{\mathrm{goal}})'Q(\boldsymbol{x}(t)-\boldsymbol{x}_{\mathrm{goal}}) + \boldsymbol{u}(t)' R \boldsymbol{u}(t),\] and the terminal cost as \[c_f = (\boldsymbol{x}(K)-\boldsymbol{x}_{\mathrm{goal}})'Q_f(\boldsymbol{x}(K)-\boldsymbol{x}_{\mathrm{goal}})\] where $Q = \text{diag}(300, 300, 500, 10, 10, 10)$, $Q_f = 2Q$ and $R = \text{diag}(10^{-3}, 10^{-3})$ are chosen positive definite diagonal weight matrices.

\begin{figure}[ht]
    \centering
    \includegraphics[width=0.46\textwidth]{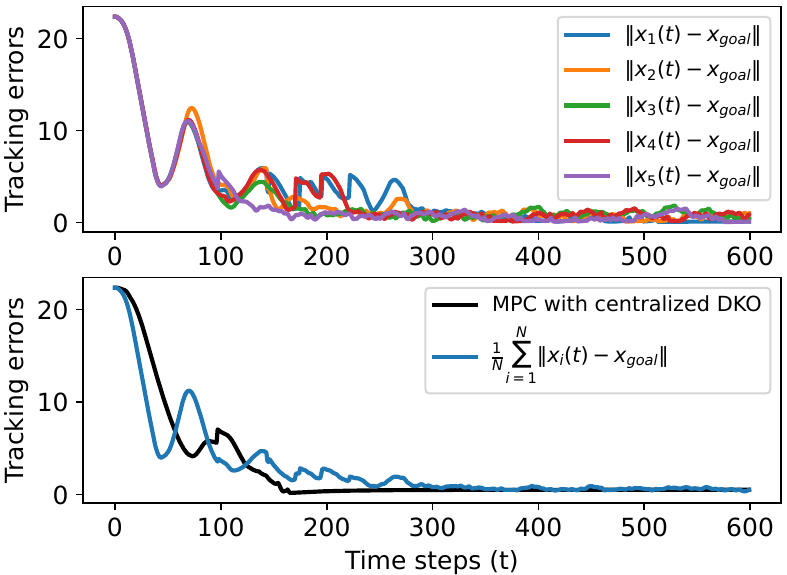}
    \caption{Tracking errors of each agent in the MAS driven by DDKL-PT-MPC}
    \label{fig:track_err}
\end{figure}

\textbf{Results Analysis.} Simulation results are shown in Fig. \ref{fig:track_err}.
By applying MPC with the deep Koopman model learned via DDKL-PT, each agent $i$ successfully reaches the goal state. The convergence time is approximately $300$ time steps across agents. Compared with MPC based on centralized DKO dynamics, the proposed method exhibits a slower average convergence rate and larger tracking errors before reaching the goal state.

\section{Conclusions}\label{conclusion}
This paper presented a distributed deep Koopman learning algorithm (DDKL-PT) for approximating the dynamics of nonlinear time-invariant systems in a multi-agent setting. Each agent learns the unknown dynamics from partial trajectory segments and exchanges its locally estimated dynamics with neighbors to achieve consensus, thereby preserving data privacy without sharing raw training data. By integrating the learned Koopman dynamics with a kinematic model, a model predictive control (MPC) framework was developed. Simulation results on a surface vehicle demonstrated both the accuracy of the distributedly learned dynamics and the effectiveness of the proposed MPC for goal-tracking and station-keeping tasks. All agents reached the desired goal state within approximately $300$ time steps, confirming DDKL-PT's capability for distributed learning and model-based optimal control.



\bibliographystyle{unsrt}
\bibliography{hao_refs}

\end{document}